\documentclass[12pt,preprint]{aastex}
\usepackage{emulateapj5,apjfonts}
\usepackage[obeyspaces]{url}
\usepackage{graphics}
\usepackage{amssymb}
\usepackage{bbm}
\usepackage{amsmath,textcomp,array}
\usepackage{onecolfloat}
\usepackage{bm}

\def\websitename{HACC Simulation Data Portal}

\lefthead{Heitmann et al.}
\righthead{HACC Cosmological Simulations: First Data Release}

\begin{document}

\def\head{

\title{HACC Cosmological Simulations: First Data Release}
\author{Katrin Heitmann\altaffilmark{1}, Thomas
  D. Uram\altaffilmark{2}, Hal Finkel\altaffilmark{2}, Nicholas
  Frontiere\altaffilmark{1,3}, Salman Habib\altaffilmark{1,4}, Adrian
  Pope\altaffilmark{4}, Esteban Rangel\altaffilmark{2}, Joseph
  Hollowed\altaffilmark{1}, Danila Korytov\altaffilmark{1,3}, Patricia
  Larsen\altaffilmark{1}, Benjamin S. Allen\altaffilmark{2}, Kyle
  Chard\altaffilmark{5}, Ian Foster\altaffilmark{5}}

\affil{$^1$ High Energy Physics Division,  Argonne National Laboratory, Lemont, IL 60439, USA}

\affil{$^2$ Argonne Leadership Computing Facility,  Argonne National Laboratory, Lemont, IL 60439, USA}

\affil{$^3$ Department of Physics, University of Chicago, Chicago, IL 60637, USA, USA}

\affil{$^4$ Computational Science Division,  Argonne National Laboratory, Lemont, IL 60439, USA}

\affil{$^5$ Data Science and Learning Division,  Argonne National Laboratory, Lemont, IL 60439, USA}

\date{today}

\begin{abstract}

We describe the first major public data release from cosmological
simulations carried out with Argonne's HACC code. This initial release
covers a range of datasets from large gravity-only simulations. The
data products include halo information for multiple redshifts,
downsampled particles, and lightcone outputs. We provide data from
two very large $\Lambda$CDM simulations as well as beyond-$\Lambda$CDM
simulations spanning eleven $w_0-w_a$ cosmologies. Our release
platform uses Petrel, a research data service, located at the Argonne
Leadership Computing Facility. Petrel offers fast data transfer
mechanisms and authentication via Globus, enabling simple and
efficient access to stored datasets. Easy browsing of the available
data products is provided via a web portal that allows the user to
navigate simulation products efficiently. The data hub will be
extended by adding more types of data products and by enabling
computational capabilities to allow direct interactions with
simulation results.
\end{abstract}

\keywords{methods: N-body ---
          cosmology: large-scale structure of the universe}}


\twocolumn[\head]

\section{Introduction}

Simulations play important roles in cosmology in many ways. They allow
us to investigate new fundamental physics ideas, to develop, test, and
evaluate new cosmological probes, and to aid large-scale surveys in
their data analysis work, pipeline development, and the investigation
of systematic uncertainties. Given the scale of the surveys and the
demands on high-accuracy predictions, these simulations are
computationally expensive and often need to be run on some of the
largest supercomputers available.

Considering the large space of analysis opportunities for the
simulations and the resource requirements needed to generate them, it
is most beneficial to share the resulting datasets within the larger
cosmology community. This need has been recognized for a long time,
and there have been multiple releases of state-of-the-art simulation
results over time, such as the Hubble Volume
simulations~\citep{hubblev}, the Millennium
simulation~\citep{springel05} and follow-on projects (e.g., a
lightcone catalog from the Millennium-XXL simulation,
\citealt{smith17}), the Dark Sky Simulations~\citep{skillman14}, and
the CosmoHub\footnote{https://cosmohub.pic.es/home} and
CosmoSim\footnote{https://www.cosmosim.org/} databases, to name just a
few -- these have had very significant impacts on the field. Following
this tradition, we describe here the first data release from a set of
simulations that will expand the currently available data sets in new
directions.

Over the last few years, we have carried out major simulations with
HACC, the Hardware/Hybrid Accelerated Cosmology Code.  HACC is a
high-performance cosmology code that has been written to enable
simulations on all available supercomputing architectures at full
system scale. Currently, HACC runs on standard X86 architectures, IBM
BG/Q machines, GPU-enhanced systems, and on Intel KNL-based
systems. Details about the code structure and performance examples are
given in~\cite{habib14}. More recently, we have added hydrodynamic
capabilities to HACC; the first results are described in
\cite{emberson18}. For this initial HACC simulation data release we
focus solely on gravity-only simulations. We have chosen simulations
that have some unique features with regard to size and the
cosmological model space covered compared to other simulations
released in the past. In particular, we release data products from the
Outer Rim simulation, described in~\cite{heitmann18}, which covers a
volume of $(4225 {\rm Mpc})^3$ and evolves more than 1 trillion
particles, and the Q Continuum simulation, described in
\cite{heitmann15}, covering a volume of $(1300{\rm Mpc})^3$ and
evolving more than 0.5 trillion particles. At their respective mass
resolutions, these are among the largest cosmological simulations
carried out to date. In addition, we release a first set of dataset
suites from the Mira-Titan Universe runs, large simulations that cover
a range of cosmologies beyond $\Lambda$CDM. This set of simulations
can be used in many ways, from generating synthetic catalogs for large
surveys, to detailed structure formation studies down to small spatial
scales, to understanding the effects of cosmological parameters on a
range of cosmological probes targeted by current and future surveys.

For our release, we take advantage of
Petrel\footnote{https://petrel.alcf.anl.gov}, a data
service pilot program hosted at the Argonne Leadership Computing
Facility (ALCF). Petrel combines storage hardware and
Globus\footnote{https://www.globus.org/} transfer
capabilities~\citep{globus} to enable easy and scalable management and
sharing of large datasets. In order to access our data, users simply
need to authenticate using a Globus-supported identity, after which
they can download data using Globus transfer capabilities.

The paper is organized as follows. In Section~\ref{sec:prod} we
describe the simulations that are part of this first data release. In
Section~\ref{sec:infra} we introduce the data portal infrastructure
and interface. Section~\ref{sec:access} provides a brief overview
regarding data access at our webportal\footnote{https://cosmology.alcf.anl.gov/}. We conclude in Section~\ref{sec:conc} and
provide an outlook on planned developments.

\begin{figure*}[t]
\centering\includegraphics[width=7in]{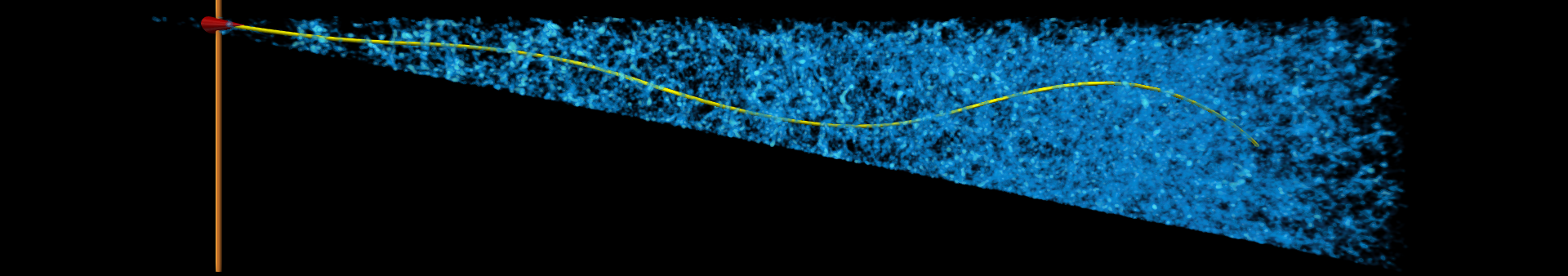}
\caption{Lightcone visualization of a small part of the SPHEREx sky
  (angular size: 130 sq degrees, redshift depth: $z=1.4$) based on the
  halos from the Outer Rim simulation.}
\label{fig:lightc}
\end{figure*}

\section{Data Products}
\label{sec:prod}
In the following we describe the data products being made available to
the community. Generally, HACC simulations produce a range of outputs
from the raw particle data to halo catalogs and merger
trees. Value-added products such as galaxy catalogs and lensing maps
can be generated from these simulation outputs. We plan to deliver
these value-added products in the future as they become publicly
available.

Our data products are classified via three categories. Level~1 data
sets represent the lowest level of output data from the simulations,
such as the raw particle outputs, and comprise the largest amount of
data. Given the size of the raw outputs (up to several PBytes for our
largest simulations) it is very difficult (and expensive) to provide
access to Level~1 data. Level~2 data has an added layer of processing,
which makes it roughly an order of magnitude smaller in size; examples
include downsampled particle snapshots, halo catalogs, and halo merger
trees. Most scientific results are based on the analysis of Level~2
data and post-processing thereof to generate Level~3 data. Level~3
data includes maps and object catalogs at different wavelengths. In
this first data release, we focus on Level~2 data, namely downsampled
particle outputs and halo catalogs. In future releases, we plan to add
Level~3 data and additional Level 2 data products.

The first public data release contains results from two very large
$\Lambda$CDM simulations, the Outer Rim simulation~\citep{heitmann18}
and the Q Continuum simulation~\citep{heitmann15} and 11 medium-sized
simulations covering varying dark energy equation of state models from
the Mira-Titan Universe suite described in~\cite{heitmann16} and
\cite{lawrence17} (the full suite contains an excess of 110
simulations).

\subsection{The $\Lambda$CDM Universe}

We begin by describing the two large $\Lambda$CDM simulations run with
identical cosmological parameters. The chosen parameter values are
close to the best-fit WMAP-7~\citep{wmap7} cosmology, given by
$\omega_{\rm cdm}=0.1109$, $\omega_{\rm b}=0.02258$, $n_s=0.963$,
$h=0.71$, $\sigma_8=0.8$, and $w_0=-1.0$. In the following, we provide
details about the size of the simulations, the associated mass and
force resolution, and the released data products.

\subsubsection{The Outer Rim Simulation}

The Outer Rim simulation, run some years ago, is still one of the
largest simulations at its mass resolution currently available. It was
carried out on Mira, an IBM BG/Q system at the ALCF. The simulation covers a volume of $(4225 {\rm
  Mpc})^3$ and evolved 10240$^3$ particles, leading to a mass
resolution of $\sim 2.6\cdot 10^9$M$_\odot$. (Note that while we quote
the volume here in units of Mpc and M$\odot$, the data itself is
delivered in units of $h^{-1}$Mpc and $h^{-1}$M$_\odot$.) The force
resolution is approximately 4kpc. (The Euclid Flagship simulation is
similar, with slightly lower mass resolution but increased volume.)
The Outer Rim simulation was designed to be most useful for answering
survey related questions and serve as the basis for generating
detailed synthetic sky catalogs. The large volume enables the
construction of large-area synthetic sky maps; at the same time the
simulation has good enough resolution to capture halos that host
galaxies targeted by the surveys, such as luminous red galaxies (LRGs)
and emission line galaxies (ELGs).

Outer Rim simulation results have been used for a number of projects:
halo catalogs for quasar clustering studies with eBOSS
\citep{eboss1,eboss2,eboss3}, a detailed investigation of the
concentration--mass relation~\citep{child18}, and strong lensing
investigations~\citep{li18}. The Outer Rim run is currently being used
to create a major synthetic sky catalog for the second data challenge
(DC2) carried out by the Large Synoptic Survey Telescope Dark Energy
Science Collaboration (LSST DESC)~\citep{cosmoDC2}. 
Scientific results, some implementation details of HACC on
Mira, the IBM BG/Q architecture, and a list of the stored outputs are
given in \cite{heitmann18}.

In future gravity-only simulations, we aim to improve the mass
resolution by an order of magnitude (similar to the Q Continuum mass
resolution) by increasing the number of particles while keeping the
volume the same or increasing it slightly.

For the current data release, described here, we provide the following
outputs at discrete time snapshots:
\begin{itemize}
\item Friends-of-friends halo properties (linking length $b=0.168$) --
  the number of particles in the halo, a halo tag, the halo mass in
  units of $h^{-1}$M$_\odot$, the potential minimum halo center
  $(x,y,z)$ in $h^{-1}$Mpc, the center of mass $(x,y,z)$ in comoving
  $h^{-1}$Mpc, the mean velocity $(v_x,v_y,v_z)$ in peculiar comoving
  km/s, and the halo velocity dispersion. Note that the halo tag is
  not consistent across time steps\footnote{Full merger trees would have to be constructed in order to provide this information. We will provide merger trees in the future, but they are not part of this first release. }. The smallest halo contains 20
  particles and therefore has a mass of $\sim 5.2\cdot
  10^{10}$M$_\odot$.
\item 1\% of all particles in a simulation snapshot, randomly selected
  (positions in comoving $h^{-1}$Mpc and velocities in peculiar
  comoving km/s).
\end{itemize}
The snapshots are stored at nine redshifts:
$z=\{1.494,1.433,\\0.865,0.779,$
$0.539,0.502,0.212,0.050,0.000\}$. The halos were identified with a
standard Friends-of-Friends (FOF) halo finder \citep{davis85} with a
linking length of $b=0.168$, a commonly used value for generation of
synthetic galaxy catalogs.

In addition, we provide lightcone outputs in one octant of the sky for
\begin{itemize}
    \item Full particle information (positions, velocities, and scale factor) out to redshift $z\sim2.3$.
    \item FOF halo properties (positions, velocities, scale factor, and mass) out to redshift $z\sim 3$.
\end{itemize}
All units are the same as for the snapshot data. The halo lightcones
are created from 57 discrete time snapshots, while the particle
lightcones are based on 49 snapshots. For the particle lightcone,
particles with the same ID are identified in two adjacent snaphots and
then the position of the particle where it intersects the lightcone is
determined by linear interpolation. The volume of the Outer Rim
simulation is not quite large enough to cover the full lightcone out
to $z=3$. We therefore employ standard replication procedures to
enable the construction of a full octant out to $z=3$. (For a detailed description of our approach, see~\citealt{hollowed19,cosmoDC2} and references therein.) For the halo
lightcone, we use detailed merger trees to identify a halo's
progenitor and then carry out an interpolation. A very detailed
discussion of the procedure can be found in \cite{hollowed19,cosmoDC2}.  As one example data product generated from the halo
lightcone data, Figure~\ref{fig:lightc} shows a small part of the
SPHEREx sky \citep{spherex} which is populated with galaxies using a
halo occupation distribution method that takes into account different
types of galaxies as seen by SPHEREx, a proposed all-sky near-infrared
spectral survey. A fly-through movie is available on
YouTube\footnote{https://www.youtube.com/watch?v=I0D7\_0Kus8g}.

\subsubsection{The Q Continuum Simulation}

The Q Continuum simulation was carried out on Titan at the Oak Ridge
Leadership Computing Facility, taking advantage of Titan's GPU/CPU
architecture. Selected science results and implementation details for
HACC on CPU/GPU platforms are presented in~\cite{heitmann15}.  The
simulation covers a smaller volume compared to the Outer Rim
simulation ($(1300 {\rm Mpc})^3$) but provides better mass resolution
by more than an order of magnitude at $\sim 1.48\cdot 10^8$M$_\odot$
(the force resolution is 3~kpc). The simulation evolved 8192$^3$
particles and was designed to enable detailed structure formation
studies at high mass resolution and very good statistics. In addition,
while the volume is not large enough to enable the construction of
large-scale sky survey maps, it offers a very valuable test bed for new
modeling approaches to investigate the galaxy-halo connection. In the
past, the Bolshoi simulation~\citep{klypin11} has been used very
successfully for this purpose but due to the smaller box size of $(250
h^{-1}{\rm Mpc})^3$, has limits with regard to available halo
statistics at higher masses. In~\cite{child18} we present results for
the concentration--mass relation from the Q Continuum simulation,
taking advantage of the large number of halos available in different
mass ranges at high mass resolution.

\begin{figure}[t]
\includegraphics[width=3.5in]{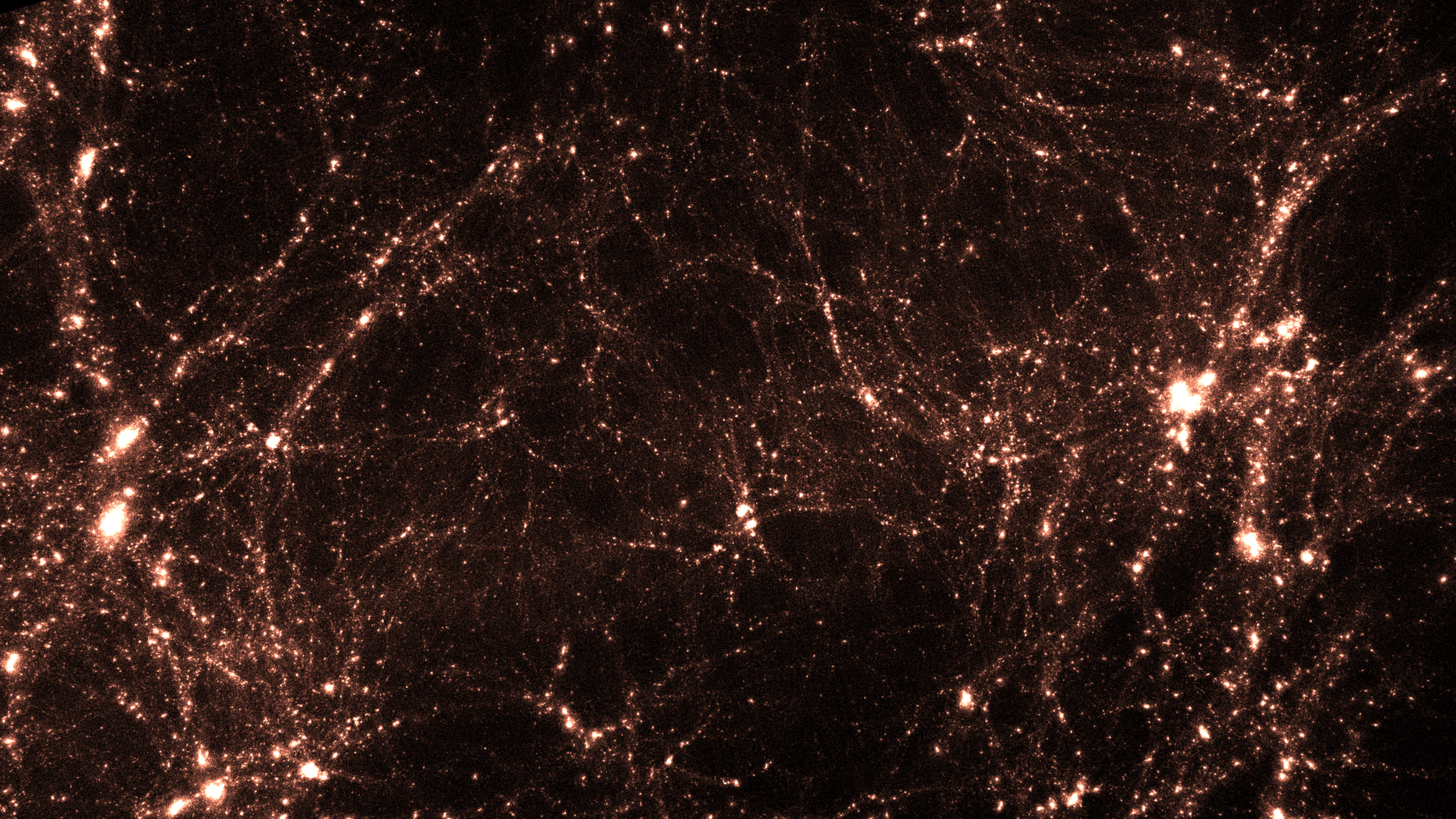}
\caption{Zoomed-in view of the particle distribution at $z=0$ from one
  MPI rank of the Q Continuum simulation.}
\label{fig:qc}
\end{figure}

Figure~\ref{fig:qc} shows a zoomed-in view of the particle distribution
from one MPI rank (the simulation was carried out using a total of
16,384 ranks) from the Q Continuum simulation at the final redshift
$z=0$. As in the case of the Outer Rim simulation, we release FOF
halos at nine redshifts (the same redshifts were chosen for both simulations) and downsampled particle snapshots (1\% randomly sampled as for the Outer Rim simulation). The
smallest halos contain 40 particles and therefore have a mass of $\sim
5.92\cdot 10^9$M$_\odot$.

\subsection{Beyond $\Lambda$CDM: The Mira-Titan Universe}

\begin{figure}[t]
\includegraphics[width=3.5in]{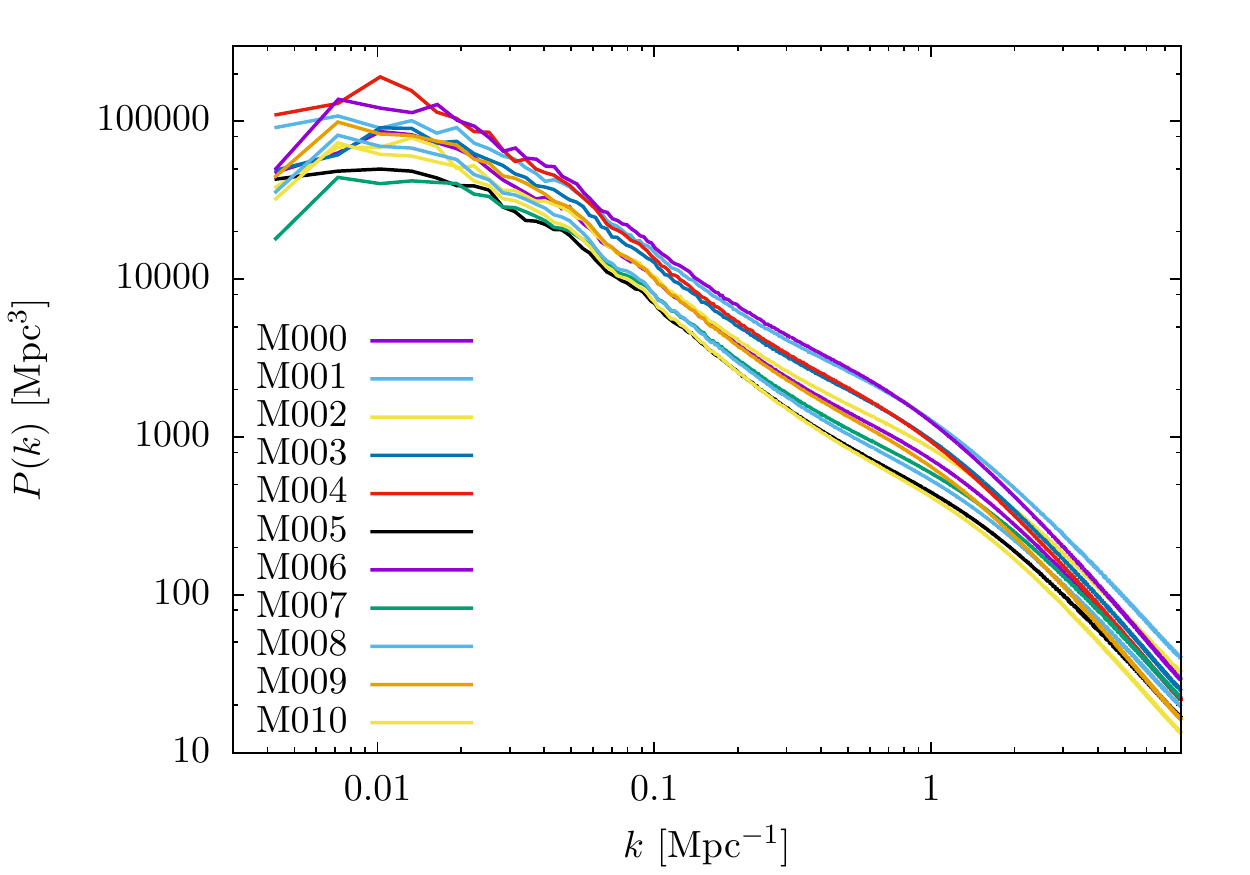}
\caption{Power spectra at $z=0$ for the different Mira-Titan
  cosmologies. M007 - M010 have the same seed, allowing for a direct
  comparison of the evolution and properties of specific structures
  across cosmological models.}
\label{fig:pks}
\end{figure}

The Mira-Titan Universe suite -- named after the two supercomputers it
was generated on -- is a unique set of simulations that spans a wide
range of cosmological models. In the first public release we include
data from 11 models that vary 7 parameters, namely $\theta=\{\omega_m,
\omega_b, h, \sigma_8, n_s, w_0, w_a\}$. All of these models and their
parameters are listed in Table~\ref{tab:models}. Each simulation
covers a $(2100$Mpc)$^3$ volume and evolves 3200$^3$
particles. Figure~\ref{fig:pks} shows the results for the nonlinear
density fluctuation power spectra at $z=0$ for all 11 models.

The complete Mira-Titan Universe suite also includes models with
massive neutrinos, and results from these will be released in the near
future. The suite encompasses 111 models and has been designed with a
tessellation-based sampling strategy to build a range of cosmological
emulators with known convergence properties. The model selection of
the subset presented here is based on a symmetric Latin-Hypercube
(SLH) design (for details on SLH designs in the context of
cosmological emulators see~\citealt{coyote2}) for M001 -- M010
covering the following parameter ranges:
\begin{eqnarray}
0.12\le &\omega_m& \le 0.155,\\
0.0215\le &\omega_b& \le 0.0235,\\
0.7\le &\sigma_8& \le 0.9,\\
0.55\le &h& \le 0.85,\\
0.85\le &n_s& \le 1.05,\\
-1.3\le &w_0& \le -0.7,\\
-1.73\le &w_a& \le 1.28.
\end{eqnarray}

The choices for these ranges are discussed in detail in
\cite{heitmann16}. For the parameterization of the dark energy
equation of state, we follow the common definition introduced
by~\citep{chevalier,linder}: $w(a)=w_0+w_a(1-a)$, where $a=1/(1+z)$ is
the expansion factor. We also provide results for a $\Lambda$CDM
model, M000, using the same cosmological parameters as for the Outer
Rim and Q Continuum simulations. Model M000 has been used in the past
for a few science investigations already, including the generation of
a galaxy power spectrum emulator~\citep{kwan15}, simulations of the
pairwise kinematic Sunyaev--Zel'dovich signal~\citep{flender16}, and in
the first Mira-Titan Universe paper to study requirements on the
simulation volume and resolution to enable precision emulation of the
matter power spectrum~\citep{heitmann16}.

\begin{table*}[t]
\begin{center}
\caption{Cosmological models \label{tab:models}}
\begin{tabular}{ccccccccl}
Model &	$\Omega_m$	& $\omega_b$&	$h$	& $\sigma_8$ &	$n_s$ &	$w_0$&	$w_a$ & Seed\\
\hline\hline
M000 & 0.2200 &	0.02258 &	0.7100  &	0.8000  &	0.9630  &	-1.0000  &	0.0000 & 5126873-M\\
	M001  &	0.3276 	 &0.02261 & 	0.6167  &	0.8778  &	0.9611  &	-0.7000  &	0.6722 &  9009097-M\\
	M002  &	0.1997 	 &0.02328 &	0.7500  &	0.8556  &	1.0500 	 &-1.0330  &	0.9111 & 29128379-M\\
	M003  &	0.2590 	 &0.02194 	 & 	0.7167  &	0.9000  &	0.8944  &	-1.1000  &	-0.2833 & 5988728-M\\
	M004  &	0.2971  &	0.02283  & 	0.5833  &	0.7889  &	0.8722 	 &-1.1670  &	1.1500 & 82983990-M\\
	M005  &	0.1658 	 &0.02350  & 	0.8500  &	0.7667  &	0.9833  &	-1.2330  &	-0.0445 & 3029384-M\\
	M006  &	0.3643 	 &0.02150  & 	0.5500  &	0.8333  &	0.9167 	 &-0.7667  &	0.1944 & 2390480-M\\
	M007  &	0.1933 	 &0.02217  & 	0.8167  &	0.8111  &	1.0280  &	-0.8333  &	-1.0000 & 5126873-T\\
	M008  &	0.2076 	 &0.02306  & 	0.6833  &	0.7000  &	1.0060  &	-0.9000  &	0.4333 & 5126873-T\\
	M009  &	0.2785 	 &0.02172  &	0.6500  &	0.7444  &	0.8500  &	-0.9667  &	-0.7611 & 5126873-T\\
	M010  &	0.1718 	 &0.02239  &	0.7833  &	0.7222  &	0.9389  &	-1.3000  &	-0.5222 & 5126873-T\\
\end{tabular}
\end{center}
\end{table*}

Since the data sets from the Mira-Titan Universe are smaller compared
to the Outer Rim and Q Continuum simulations we release a broader set
of outputs. We provide, for each of 27 redshifts,
$z$=\{4.000,3.046,2.478,2.018, 1.799, 1.610, 1.376, 1.209, 1.006,
0.779, 0.736, 0.695, 0.656, 0.618, 0.578, 0.539, 0.502, 0.471,
0.434,0.402, 0.364, 0.304, 0.242, 0.212, 0.154, 0.101, 0.000\}, the
following data products:

\begin{itemize}
\item Friends-of-friends halo properties: the same halo properties
  as for the Outer Rim simulation are provided, down to 20 particles
  per halo, halo masses depend on the exact cosmology but start at
  $\sim 10^{11}$M$_\odot$.
\item Downsampled halo particles: for each halo we provide the
  positions and velocities for 1\% of the halo particles, randomly
  chosen. If the halo size is smaller than 500 particles, we provide 5
  particles instead of 1\%.
\item Full halo particles: for halos that have at least 1000
  particles, we provide positions and velocities of all halo
  particles. This information overlaps with the downsampled particles
  but both catalogs can be used for different purposes and we
  therefore wanted to provide them both as a complete, independent
  set.
\item 1\% of all particles in a simulation snapshot, randomly selected
  (positions in comoving $h^{-1}$Mpc and velocities in peculiar
  comoving km/s).
\end{itemize}

The last column in Table~\ref{tab:models} indicates the machine the
simulation was run on (M: Mira, T: Titan) and the random seed for
setting the initial condition. If the seed and the supercomputer are
the same (as they are for M007-M010) the simulations have the same
phases in the initial condition and can be compared directly,
structure by structure, instead of just statistically via, e.g., the
power spectra or mass functions. This feature is of potential interest
for some comparative investigations.

\section{Data Portal Infrastructure}
\label{sec:infra}
The very large HACC simulations that we describe in this paper
generated petabytes of raw data. During production, the data reside on
the file systems attached to the supercomputers on which the
simulations are run: multiple machines at two facilities over the
course of multiple allocation years. As with compute hours, this
dedicated storage expires with the allocation year. To support the
publication of this data set, we have aggregated a subset of the data
to a common, shared storage infrastructure, Petrel, at the ALCF. An
accompanying web portal simplifies the task of selecting and
transferring data from Petrel to a user's institutional compute
resources via Globus. (A pre-requisite to browse and access our data
is authentication via a Globus-supported identity.) The
GenericIO-formatted files can be read with the GenericIO
library\footnote{https://trac.alcf.anl.gov/projects/genericio}, which
is open source and available from the portal. The GenericIO library
provides a C++ code for reading and writing GenericIO-formatted data
in parallel and a Python wrapper for reading data serially (suitable
for smaller data). The following sections describe the portal,
storage, and file format in more detail.

\subsection{Petrel}

Petrel is a flexible data service designed to support collaborative
access to large datasets within scientific communities. Building on a
high-performance 1.7PB parallel file system (to be increased in the
near future) and embedded in Argonne National Laboratory's 100+ Gbps
network fabric, Petrel leverages Science DMZ concepts and Globus APIs
to provide high-speed, highly connected, and programmatically
controllable data management.

Petrel provides a user-oriented, collaborative storage model in which
users manage their own isolated storage allocations. They are able to
upload data to their allocation, manage the organization of data, and
also dynamically share data without requiring local user accounts.  It
provides both Web and API access, via Globus, which allow
sophisticated applications, such as the data portal described here, to
be implemented with modest amounts of programming.

Petrel overcomes two important challenges associated with managing and
using large data: high-speed data access and flexible collaborative
usage.  These aims have long been at odds with the way data is made
accessible at high-performance computing (HPC) centers.  For example,
data is typically stored either on parallel file systems designed for
rapid internal access or on specialized data portal servers that
support slower external data distribution. Thus high-speed data
movement in and out of HPC centers is often difficult.  Secondly,
dynamic collaboration is challenging due to the need for individual
user accounts and policies regarding who may create accounts. In many
cases the process for obtaining accounts lacks the flexibility
required to support dynamic collaborations.

Petrel is designed to support implementations of the modern research
data portal (MRDP) design pattern~\citep{mrdp}, in which control
channel communications and data channel communications are separated,
with the former handled by a web server computer deployed (most often)
in the institution's enterprise network and the latter by specialized
data servers connected directly to high-speed networks and storage
systems.  This pattern has emerged due to the availability of two
technologies. The first, the Science DMZ~\citep{dart13}, enables
high-speed data access by connecting specialized data servers directly
to high-speed networks and storage systems. The second, Globus,
enables the outsourcing of crucial data portal functionality such as
managing data transfers, data access, and authentication.

\subsubsection{Petrel Data Store and DTNs}

The Petrel system comprises a parallel file system with eight
specialized data transfer nodes (DTNs)~\citep{dart13} for fast remote
access.  It is configured to operate as a Science DMZ, enabling access
from external networks without passing through the usual
firewalls. Line-rate firewalls are in place, configured with network
access control lists (ACLs) to limit access to Globus/GridFTP ports, and in
particular, to limit control channel access to Globus service IP
addresses.

The eight Petrel DTNs are connected to two core Mellanox SX1710
36-port 40GbE switches maintained within Argonne's Joint Laboratory
for Systems Evaluation (JLSE). The Petrel DTNs are split across the
two switches, each connected with one 40GbE. Each of the two 40GbE
core switches has a 2 $\times$ 40GbE link aggregation group to the ALCF
core router, which in turn has a 100GbE connection to the core router
for the data center within which the ALCF is located, which in turn
connects at 100GbE to one of the Argonne border routers.  Thus, as
Petrel traffic reaches beyond JLSE to ALCF, the data center, and the
border, it shares bandwidth with an increasing range of other
activities.

\subsubsection{Globus Remote Access Services}
Petrel relies on Globus~\citep{globus} for identity and access
management (IAM), data management, and high-speed data transfer. These
capabilities are delivered as a cloud-hosted services, enabling users
to access them through their web browser and developers to invoke them
via REST APIs.

Petrel uses Globus Auth~\citep{tuecke2016} to provide IAM
capabilities. This allows users to manage and access data on Petrel by
authenticating using a Globus-supported identity (e.g., institution,
ORCID, or Google accounts).  It also allows developers to securely
access Petrel via Globus APIs by using OAuth~2.0 to obtain access
tokens (on behalf of users or other services) for performing actions.

Each allocation on Petrel is underpinned by a Globus shared
endpoint~\citep{globussharing}---a virtual construct that allows for
management of access permissions within a particular directory of the
host (Petrel) endpoint. The owner of the allocation is granted
administrator privileges on the shared endpoint, allowing them to
associate an ACL with files or folders in the
endpoint. ACLs, including read and write permissions, may be granted
to specific users or groups of users.  Shared endpoints operate
entirely with Globus-supported identities, thus neither
administrators nor users of a shared endpoint need to have local
accounts.

Users are granted complete control of their allocation, enabling them
to transfer or upload data into the allocation, manage data within the
allocation (e.g., create directories, rename files, delete files),
download data using secure HTTPS, and transfer data to other Globus
endpoints.  All such functions can be performed by using the Globus
web interface or via JSON-based REST APIs.

\begin{figure}[b]
\centering\includegraphics[width=3.5in]{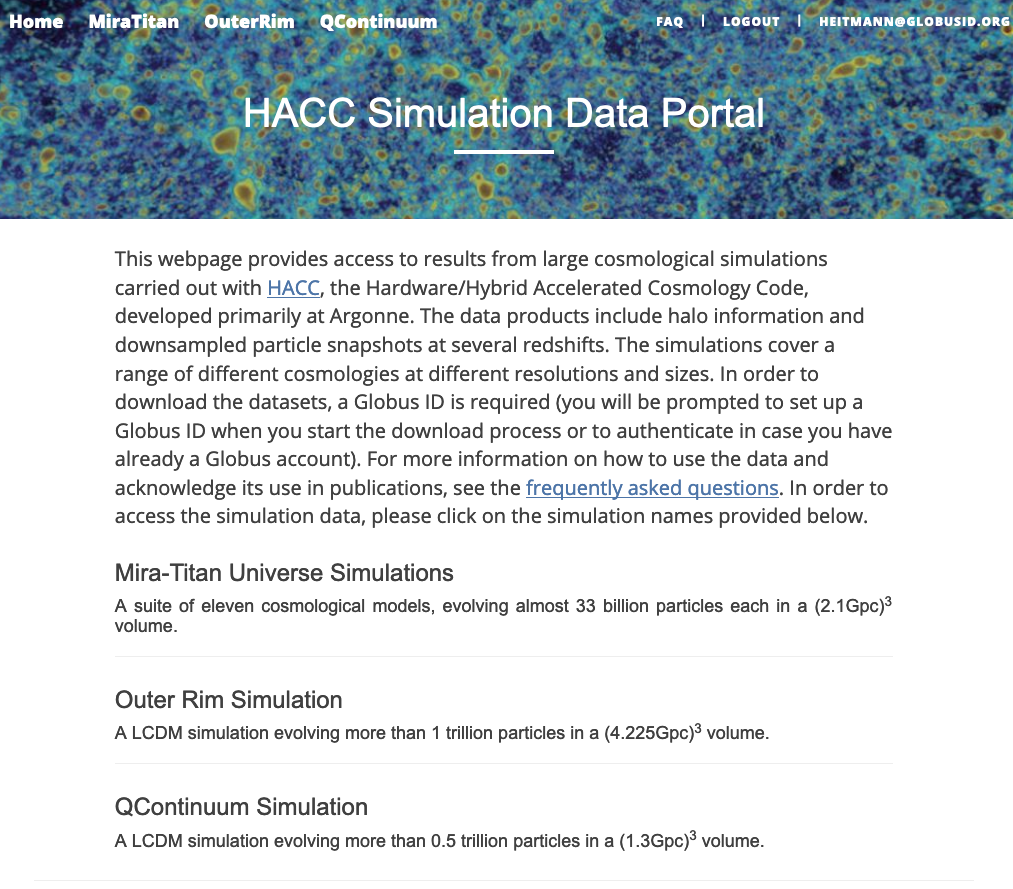}
\caption{Image of the main {\websitename} page that describes the
  available simulations.}
\label{fig:portal_main}
\end{figure}

\subsection{The Web Portal}
\begin{figure}[t]
\centering\includegraphics[width=3.2in]{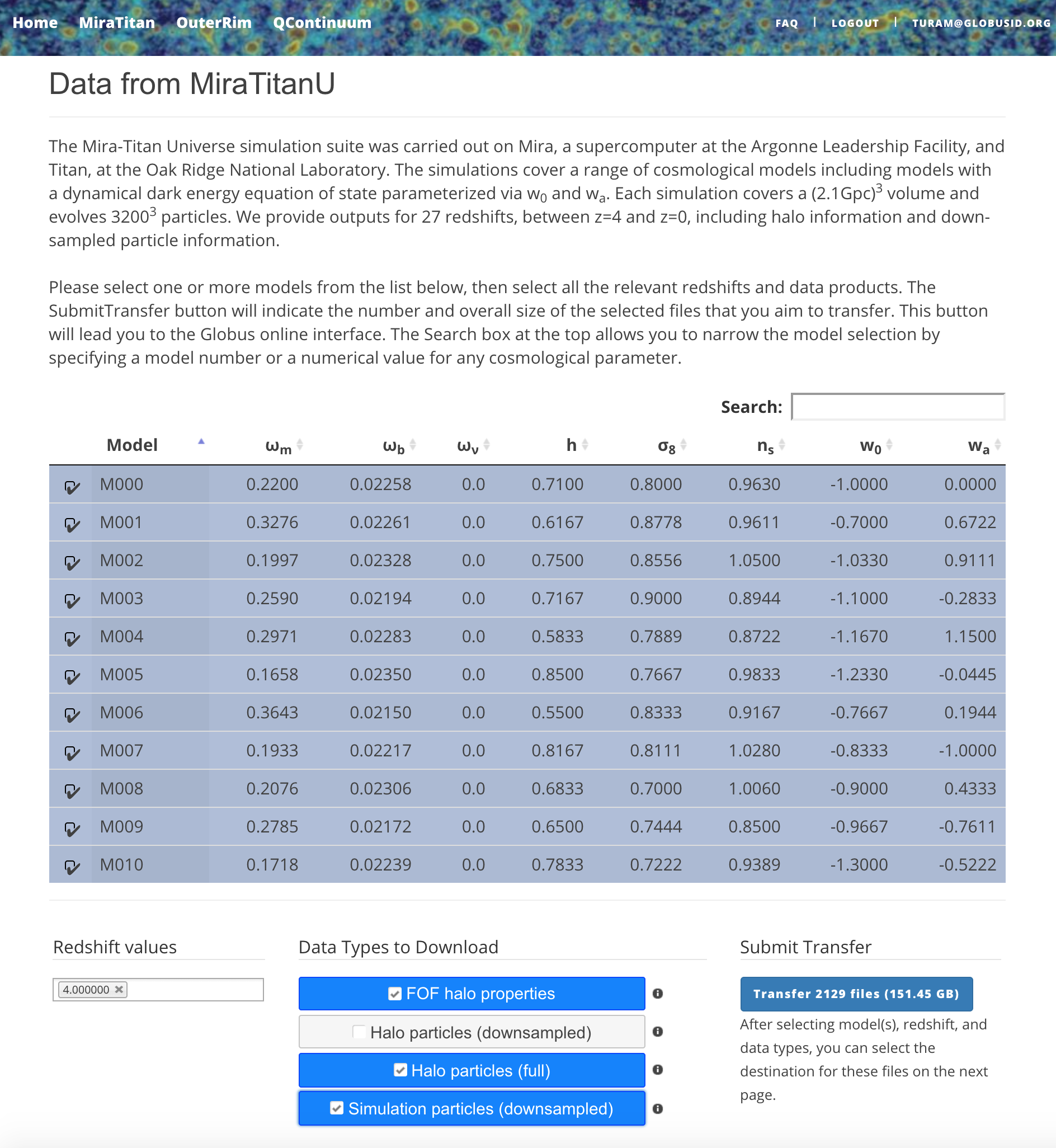}
\caption{Snapshot of the transfer page where the user can select
  simulation data to download, by choosing cosmological model(s),
  redshift(s), and data type(s). Information about data sizes is
  provided after the datasets have been selected.}
\label{fig:portal_select}
\end{figure}

The {\websitename} (shown in Figures~\ref{fig:portal_main} and
\ref{fig:portal_select}) offers users an easy means of selecting
simulation data and beginning a transfer to their home institutions
for further analysis.  This functionality is essential given that
users are typically interested in analyzing particular subsets of the
data. The portal permits selection of subsets of the data by
simulation suite, model, redshift, and data type. In the case of the
Mira-Titan Universe simulations, data is available from models
simulated with eleven different sets of cosmological parameters, as
enumerated in Table~\ref{tab:models}. Users can select a range of
associated redshift values, and the data types of interest, and
proceed to Globus, where they can select a destination for the
transfer.

Users can browse data on the portal without authenticating. Transfers
are managed by Globus: after selecting data in the {\websitename}, the
user is redirected to Globus to select a destination endpoint:
typically, the user's home institution, but the user could, of course,
instead transfer the data to any compute resource for analysis. Once
the transfer has been submitted, the user is redirected back to a
transfer status page on the {\websitename} server. Transfer activity
can also be viewed on Globus.

In the example given in Figure~\ref{fig:portal_select}, the user has
chosen to download halo properties, full sets of halo particles, and
downsampled simulation particles from eleven models, with one redshift
value, resulting in a transfer of 2129 files, ranging in size from 4KB
to 794MB (total 151~GB); this is a manageable subset of the available
data.  As a point of reference, transfers of this dataset to the ALCF
and the National Energy Research Scientific Computing Center (NERSC)
completed in 48.2~s and 40.1~s, respectively. Smaller
transfers are, of course, possible (for example, by selecting halo
properties from a single model and a high redshift, when fewer halos
would have formed). The portal maintains a record of downloads to aid
our understanding of how the data is being used, so we can tailor
current and future data releases.

The portal is an implementation of the MRDP design pattern, which
combines a modern web front end for browsing and discovering data, with
Globus services on the server for managing data access and
movement. This component-based approach makes it possible for the
front-end portal to reside on publicly accessible server nodes to
accept user requests, while calls to Globus enable direct data
transfers from the Petrel DTNs, sitting in the Science
DMZ, to the user's destination resource.  We extended the example MRDP
by developing mechanisms for creating and updating the catalogs of
simulation data, developing functionality to enable querying and
filtering the data, and creating an intuitive web interface to make
these capabilities available to users. Our portal supports queries
based on model and cosmological parameters, and selecting slices of
the data across model and redshift values. The example MRDP was a very
useful starting point for building a powerful data transfer portal,
particularly in that the support for authentication and transfer
worked without requiring any additional effort.

\section{Data Access}
\label{sec:access}

\subsection{Directory Structure for Downloaded Data}

The setup of our {\websitename} allows the user to build a HACC
simulation directory at their home institution over time. At the first
requested data transfer, a directory structure is built with the
following layout in the user chosen root directory, which could be
called \url{HACC_Simulations}. At the top layer, the simulation suite
or name will appear (for this release, this will be either
\url{OuterRim}, \url{QContinuum}, or \url{MiraTitanU}). Next, a
directory will be created with the model name as it appears on the
web portal (e.g., \url{M000}), then a directory will appear that
indicates the box size in Mpc. After that, a directory name that
indicates the realization of the simulation will be built, e.g.,
\url{HACC000}. While in this release we have only one realization per
model, in later releases this might change. In addition, this
directory name helps to easily identify models that have been run with
the same seed. For example, in Table~\ref{tab:models}, we indicate in
the last column that the seeds for models M007-M010 in the
Mira-Universe suite are the same. This is reflected in the
\url{HACC007} directory name. Finally, a data analysis directory,
\url{analysis} is built that contains sub-directories for the halo,
particle, or lightcone outputs (for the simulations they are
available). Within those directories, more sub-directories are built
depending on the user's data choices.

When the user requests the next transfer and points Globus Transfer to
the same root directory, above \url{HACC_Simulations}, the overall
directory structure will be preserved and the new data will be added
accordingly. As a concrete example, if the user downloaded in the
first transfer \url{M000} data from MiraTitanU and now decides to
investigate \url{M002}, the \url{M002} directory will automatically
placed into the MiraTitanU directory. Another example could be that
the user only requested halo properties during the first transfer for
\url{M000} and now also requests downsampled particles. In this case,
the files will be automatically be placed into the appropriate analysis
directory for \url{M000}. This setup will help the user to keep a
clean directory structure that can be built up over time
automatically.

\subsection{GenericIO Library}

Data obtained from the {\websitename} uses the GenericIO data
format. Efficient I/O is extremely important in the context of
simulations at scale. In order to achieve optimal write speed, we
developed a customized I/O library for HACC, called GenericIO. At
simulation time, the GenericIO library, described in some detail in
\cite{habib14}, conducts a hierarchical aggregation of the outputs
from the simulation ranks to match the design of the interconnect and
I/O subsystem on a particular supercomputer. Instead of writing a
single monolithic file of the particle data, which would be very
inefficient, GenericIO subdivides the output into several files
(depending on the simulation size, $\mathcal{O}$(100)) in order to
achieve a high percentage of the peak performance of the available
bandwidth. The parameters that GenericIO uses for aggregation are
tunable to accommodate the design of diverse I/O systems and different
output sizes.  For example, on Mira, a BG/Q architecture with a GPFS
file system, GenericIO writes one file per dedicated I/O node in the
system. On that system there are 128 compute nodes for each I/O node,
leading to a 128-fold reduction in the number of output files.

The GenericIO library is open source and includes C++ and Python
interfaces for accessing the datasets obtained from the portal. In
order to build the
library\footnote{https://trac.alcf.anl.gov/projects/genericio} a
reasonably modern C/C++ compiler is required; for example, the library
builds successfully with GCC 4.8.5. Two build options are provided
with GenericIO: serial tools that can run on the front end of a
machine, including a Python library; and an MPI library that can be
linked into a parallel executable.

The front-end programs can be built with the command
\begin{verbatim}

    make frontend-progs

\end{verbatim}
This build includes two command-line executables. The GenericIOPrint
executable is run as \textit{GenericIOPrint gio\_datafile} to view the
file content as text; GenericIOVerify, executed similarly, is used to
confirm the internal consistency of GenericIO-formatted data files. An
example of building the frontend-progs target and a subsequent Python
session is given in the Appendix~\ref{append}.

The MPI programs can be built with the command
\begin{verbatim}

    make mpi-progs

\end{verbatim}
This build includes GenericIOPrint and GenericIOVerify, described
above, and three other programs: GenericIORewrite,
GenericIOBenchmarkRead, and GenericIOBenchmarkWrite.

The GenericIORewrite program is used to rewrite existing files with a
different number of MPI ranks or with a subset of the original fields:
\begin{verbatim}

GenericIORewrite gio_old gio_new \
    [field1 [field2...]]

\end{verbatim}
The benchmark codes are used for benchmarking writes and reads to
parallel file systems. The source code of the command-line tools serves
as a good reference for how one might use the GenericIO library in an
application; the benchmark codes are perhaps the most direct and
compact of these examples.

\subsection{Data Format}

As mentioned above, for each output GenericIO generates a set of files
instead of one monolithic file. Each output, which could be a snapshot
or a shell from the lightcones, will have a metadata file and a set of
subfiles that are marked with a hash. When using the reader, one can
either point it at the main, unhashed file and automatically the full
set of files will be read, or one can read each hash file
separately. In case of the snapshot files, the hash files contain
blocks of data and each block originates from a rank and holds a
contiguous region in space. However, the blocks within a hashed file
are usually not from the same spatial region so that a single file
will contain disconnected blocks that originate from different parts
of the simulation volume. If the user works with one simulation but
different data products, the regions within a hash file are the same
across the different outputs. This is sometimes convenient if, e.g.,
one inspects a large halo that was found from the FOF properties files
and wants to match the halo particles in the corresponding file.

\section{Conclusion and Outlook}
\label{sec:conc}
In this paper we have introduced the first data release from a range
of HACC simulations, including two large $\Lambda$CDM simulations as
well as a range of simulations covering different cosmologies. We use
the Petrel platform to provide convenient access to the data via
Globus. In order to access the data, the user simply needs to
authenticate using a Globus-supported identity and a Globus endpoint
to enable the data transfer. In order to make the data sets easily
searchable, we developed a web portal that allows the selections of
the desired data sets to be transferred. The sizes of the selected data
sets are summarized and the interface then directly leads the user to
the Globus interface.

The available data sets enable the exploration of different
cosmological models and can be used to build synthetic catalogs. The
web portal allows us to monitor the data transfers and therefore the
data selections. This information is very valuable for future
extensions of the service.

We plan to release additional simulations in the future and to extend our platform to allow the user to carry out limited data analysis projects directly. The Mira-Titan Universe suite overall encompasses 111 cosmological models. Depending on user requests and storage availability, we will release more models in the future. In addition, new extreme-scale simulations are already available from recent runs on the Summit supercomputer at the Oak Ridge Leadership Computing Facility. The data is currently being curated for a future release.

A more ambitious future goal is the addition of analysis capabilities
to our service. This would enable sharing and analysis of large datasets and at the same time
would allow for different levels of interactions with the data, from simple searches and
downloading capabilities, to interactive queries and processing, to enabling the
submission of larger-scale batch jobs. We recognize that many researchers will need to run further
analyses on the data, and will lack the necessary computational resources to do so
due to limitations at their home institutions (multi-petabyte storage and large-scale
computing). Accommodating these users would require additional infrastructure and
software: we have started to develop PDACS (Portal for Data Analysis Services for
Cosmological Simulations,~\citealt{pdacs}) as a means of making data available for user-facing
analysis workflows. We will continue to integrate and expand
various services within PDACS, including Jupyter
notebook deployment, Globus Online, and visualization, collaboration, and
simulation management tools into PDACS. 

\acknowledgments 
We are grateful for visualization support provided by
Joe Insley and Silvio Rizzi. We thank members of the CPAC group at
Argonne (in particular Jonas Chaves-Montero) for generating the
SPHEREx lightcone data; Andrew Hearin for continuous encouragement to
publicly release the data; and Mike Papka for setting up Petrel and
providing storage space for our simulations. We gratefully acknowledge
the computing resources provided and operated by the Joint Laboratory
for System Evaluation (JLSE) at Argonne National Laboratory.  Argonne
National Laboratory's work was supported under the U.S. Department of
Energy contract DE-AC02-06CH11357. This research used resources of the
Argonne Leadership Computing Facility, a DOE Office of Science User
Facility supported under Contract DE-AC02-06CH11357. This research
also used resources of the Oak Ridge Leadership Computing Facility, a
DOE Office of Science User Facility supported under Contract
DE-AC05-00OR22725.

\onecolumn
\appendix

\section{GenericIO Python Session}
\label{append}

In order to read the HACC data, we provide the GenericIO library. We recommend to use the C++ interface for the larger files (e.g., the particle lightcones). For smaller files, such as the halo catalogs, the Python interface provides a convenient approach to interact with the GenericIO files. Below we show an example Python session that shows how to obtain general information about the file content with {\em gio.inspect} and how to read data or a subset of the data columns with {\em gio.read}. 

\begin{verbatim}
$ git clone git://git.mcs.anl.gov/genericio.git
$ cd genericio
$ make frontend-progs
$ export PYTHONPATH=$PWD/python
$ python
>>> import genericio as gio
>>> gio.inspect('m000-99.fofproperties')
Number of Elements: 1691
[data type] Variable name
---------------------------------------------
[i 32] fof_halo_count
[i 64] fof_halo_tag
[f 32] fof_halo_mass
[f 32] fof_halo_mean_x
[f 32] fof_halo_mean_y
[f 32] fof_halo_mean_z
[f 32] fof_halo_mean_vx
[f 32] fof_halo_mean_vy
[f 32] fof_halo_mean_vz
[f 32] fof_halo_vel_disp
(i=integer,f=floating point, number bits size)
>>> fof_halo_count = gio.read('m000-99.fofproperties','fof_halo_count')
>>> print(fof_halo_count)
[[624 681  69 ... 184  72  97]]
>>> data = gio.read('m000-99.fofproperties',['fof_halo_count',
'fof_halo_tag','fof_halo_mean_x','fof_halo_mean_y','fof_halo_mean_z'])
# returns parallel arrays in input order
>>> data
array([[6.24000000e+02, 6.81000000e+02, 6.90000000e+01, ...,
        1.84000000e+02, 7.20000000e+01, 9.70000000e+01],
       [1.32871000e+05, 1.03333000e+05, 5.48230000e+04, ...,
        1.90935200e+06, 2.05578600e+06, 7.64180000e+04],
       [1.52459240e+00, 1.43878233e+00, 1.36675692e+00, ...,
        1.14827515e+02, 1.27592453e+02, 1.27921860e+02],
       [1.43614788e+01, 3.65754814e+01, 3.79349136e+01, ...,
        6.43497162e+01, 6.44614944e+01, 8.80533829e+01],
       [3.65939808e+00, 3.32679443e+01, 3.58395233e+01, ...,
        1.05952095e+02, 1.08691956e+02, 1.26013718e+02]])
\end{verbatim}

\end{document}